\documentclass[aps,prd,twocolumn,superscriptaddress,floatfix]{revtex4} 

\usepackage{graphicx,multirow}
\usepackage{amsmath}
\usepackage{enumitem}

\newcommand{\eandm}{E\&{}M}

\begin{document}
\title{Flipping the Large-Enrollment Introductory Physics Classroom}
\author{Chad T. Kishimoto}
\email{ckishimoto@sandiego.edu}
\affiliation{Department of Physics and Biophysics, University of San Diego}
\affiliation{Center for Astrophysics and Space Sciences, University of California, San Diego}
\author{Michael G. Anderson}
\email{mganders@ucr.edu}
\affiliation{Department of Physics and Astronomy, University of California, Riverside}
\author{Joe P. Salamon}
\email{jsalamon@miracosta.edu}
\affiliation{Department of Physical Science, MiraCosta College}
\date{\today}

\begin{abstract}
Most STEM students experience the introductory physics sequence in large-enrollment ($N \gtrsim 100~{\rm students}$) classrooms, led by one lecturer and supported by a few teaching assistants.  This work describes methods and principles we used to create an effective ``flipped classroom'' in large-enrollment introductory physics courses by replacing a majority of traditional lecture time with in-class student-driven activity worksheets.  In this work, we compare student learning in courses taught by the authors with the flipped classroom pedagogy versus a more traditional pedagogy.  By comparing identical questions on exams, we find significant learning gains for students in the student-centered flipped classroom compared to students in the lecturer-centered traditional classroom.  Furthermore, we find that the gender gap typically seen in the introductory physics sequence is significantly reduced in the flipped classroom.
\end{abstract}

\maketitle

\section{Introduction}

The link between student-centered active learning processes and student learning in the college physics classroom has long been adjudicated in the physics education literature.  Early work in the field showed that the traditional lecturer-centered mode of higher-education physics instruction resulted in minimal student learning in understanding the fundamental physical concepts \cite{hh85}, and that it was necessary for student-centered active learning pedagogies to be implemented in the classroom to attain significant student improvements in understanding those physical principles \cite{mcd91,hake98}. Recent work has shown that even inexperienced instructors using active-learning techniques promote higher levels of student learning than well-regarded experienced lecturers who do not \cite{dsw11}.  (See, {\it e.g.}, Ref.\ \cite{synthesis} for a synthesis of recent advances in physics education research.) 

A study across STEM disciplines showed that students learn more and fail less often in active classrooms as opposed to those in traditional lecture classrooms \cite{freeman14}. In addition, a ``gender gap'' -- the differential performance of male over female students -- in introductory physics has been observed and can persist even when active learning pedagogies are employed \cite{lcm06,pfk07,kpf09,mms13,hen17,kms18}.

Most students who receive a bachelor degree in a STEM field will encounter the introductory physics sequence in a large-enrollment classroom.   A fraction of these undergraduates have the fortune of attending a university with faculty engaged in physics education research with the availability of resources to significantly reform the large-enrollment introductory physics classroom.  The rest will experience introductory physics in a large lecture hall, with hundreds of other students, with no more than a few (often one) graduate student teaching assistants to support the course.

In this work, we describe a robust implementation of a flipped classroom while under the common large-lecture classroom constraints mentioned above.   In other words, we decided to transform our large-enrollment ($N \gtrsim 100~{\rm students}$) introductory physics classroom from a typical large-lecture active-learning environment (traditional lecture in a large lecture hall interspersed with think-pair-share clicker questions) toward one that resembles a flipped classroom.  A flipped, or inverted, classroom is one in which activities that typically take place in class, such as lectures, will take place outside the classroom, and those which typically take place outside of class, such as student-driven problem solving, will take place in the classroom \cite{lpt00}.  Although the flipped classroom is typically associated with students watching video lectures outside the classroom to make up for the lost in-class lecture \cite{bv13}, we instead provided student reading goals, suggested reading and written outlines of the material that would've been covered in a traditional lecture.  The primary goal of the transition to the flipped classroom was to replace instructor-focused lecture time with student-centered work and discussion on activity worksheets.

We will further supplement our description of this flipped implementation with an anecdotal discussion of the successes, difficulties and future goals of our implementation.  Finally, we will present an analysis of the effectiveness of this flipped classroom methodology using in-class student quizzes and final exams as a proxy for student learning, keeping course and instructor unchanged.  To compare student results to a ``control'' group with the same instructor, but without the flipped classroom reform, we used archived data.  Although this limits the ability to perform a robust set of assessments, we find that our implementation of the flipped classroom has so changed our impressions of effective instruction that we could not, in good faith, return to spending the majority of time in the classroom lecturing.  Nevertheless, we feel that the evidence we are able to assess support the efficacy of the methodology.

\section{Methodology}
\subsection{Population and Classroom Characteristics}
This study was conducted at a large, public, research-focused institution where introductory science lectures are routinely held in classrooms that hold hundreds of students.  At this institution, approximately 50\% of enrolled students were female and the other 50\% male.  The most prevalent ethnicities are Asian (47\%), Caucasian (19\%) and Mexican-American (14\%).  Incoming high school grade point averages exceed 4.0 and 28\% of students are first-generation college students \cite{ucsddata}.

We introduced the flipped classroom pedagogy in the three-quarter introductory physics sequence aimed for students in the life- and health-sciences.  Over half of these students hope to pursue professional schooling in the health sciences and these courses are $50-65\%$ women.  The three quarters of this sequence can be categorized as Newtonian mechanics (Mechanics), electricity and magnetism (\eandm), and a combination of oscillations, waves and modern physics (Waves/Modern).  The content of each course in this sequence is determined by the department and students regularly switch instructors from quarter-to-quarter.

This flipped classroom environment was created with the same constraints of other introductory physics lectures at the institution.  Each section had one instructor and one graduate student teaching assistant for a large enrollment (100-300 students) class.  Lectures met for three hours per week (over a ten week quarter) in a large lecture hall with stadium seating and fixed desks.  A discussion section met weekly, but because the sections are poorly scheduled (often late night), these sections suffer a $\lesssim 15\%$ attendance rate.  Courses utilized a learning management system to act as a course website and to administer out-of-class assignments and use in-class personal response systems (``clickers'').  Because of resource constraints, namely one teaching assistant for hundreds of students, courses administered bi-weekly multiple choice quizzes and a three-hour multiple choice final exam.  While this final point is not ideal for anyone teaching an introductory physics course, by happenstance it has made possible this analysis of the efficacy of the flipped classroom model employed.

\subsection{Flipped Classroom Methodology}

The term ``flipped classroom'' entails a variety applications of a central pedagogical structure:  the content distribution that traditionally takes place in the classroom setting is performed before class, which allows the application and individual practice of the content that traditionally takes place out of the classroom to begin in earnest in the classroom.  Just like all student-centered reforms, its success depends on both student buy-in and instructor implementation.  The former requires the persistent attention of the instructor, while the latter is affected by faculty time constraints and departmental or university-level constraints of time, space and resources. In this section, we present our implementation of the flipped classroom methodology that is feasible within the structure and limitations that are commonly relevant at institutions with large introductory physics classrooms.

The primary guiding principle in designing our flipped classroom methodology is to use learning goals as a roadmap for student preparation, in-class activities, and formative and summative assessment.  While we do not claim that this principle represents novel or revolutionary pedagogy, we have found that this principle provides structure for students in the course, facilitates student buy-in, and forces the instructor to distill the essential ideas driving the course.

The central vessel for student learning in this flipped classroom is student-centered activity worksheets.  Student-focused work on these activity worksheets comprises a majority of in-class time and most of the out-of-class resources available for students to prepare for assessments.  Using learning goals as a guide, the activity worksheets must explore every important topic and skill covered in the course at a depth appropriate for course assessments.  Worksheet exercises are scaffolded to build up important skills in the same way a well-crafted lecture would introduce these skills.  Truly, the activity worksheets represent a {\it replacement} of the traditional lecture modality:  the usual instructor processes associated with preparing to deliver a lecture are replaced by compiling worksheets that cover the content and forces students to take an active role in their learning process.

Finally, the learning goals serve as a guide to create exams that are reflective of the in-class activity worksheets.  This is necessary (a) to promote student buy-in and (b) because the worksheets address all important course topics ({\it i.e.}, if it's important, it should be in a worksheet, and if it's important, it should be assessed).

The remainder of this section will chronicle student activities before lecture, during lecture, and after lecture in preparation for an assessment.


\subsubsection{Before Lecture}

Students must enter the classroom with an appropriate level of preparation for the flipped classroom to be effective in fostering student-centered learning with minimal instructor-centered lecturing.  We addressed this by:  (1) providing learning goals to link pre-lecture preparation with in-class activity, (2) providing a set of lecture outlines sufficient to cover these goals, and (3) requiring a pre-class formative assessment.

Providing students a set of learning goals for their pre-lecture preparation allows them to focus on the important topics that will be covered in the classroom and to begin to build their understanding of and to formulate relevant questions on the subject matter.  These are the learning goals that permeate the course from preparation to in-class activities to assessment that forms the backbone of the course.

Along with suggested reading from the course textbook, we provided lecture outlines (adapted from our own lecture slides) that cover the material that concerns these learning goals.  In the context of the flipped classroom, this is typically addressed with video lectures.  We chose to focus our time and energy on other aspects of the flipped classroom rather than undergo the time-consuming process of creating video lectures.

This was a tactical decision on our part and the results of this work are in no way intended to adjudicate the efficacy or necessity of video pre-lectures in the flipped classroom methodology.  While some work has shown the effectiveness of a multimedia pre-lecture presentation with interspersed required student questions \cite{sbgm10,sad12}, the results from this work have been achieved without such activities that are time and resource-intensive on the teaching side.  Other work has shown that the pedagogical methods employed in creating pre-lecture videos are important to their utility as a learning tool so that, from a student learning perspective, the time spent creating clear, concise expositions of the pre-lecture material may be spent in vain \cite{msr08,mbsr08}.  It remains an open question whether this flipped classroom method could show additional benefits from improved pre-lecture presentation.

Of course, students need both guidance to effectively read in preparation for class and incentives to complete the assigned reading before the relevant classwork.  Learning goals and lecture outlines provide students structure to focus their pre-lecture activities in useful directions.  One of us has even taken the time to produce reading guides to direct students' attention and thought-processes in suggested reading from the course textbook.  Reading incentives for students to perform their pre-lecture preparation before class can include multiple choice reading quiz questions relevant to the learning goals \cite{pimazur,fiveknight} or free-response questions that probe the learning goals in the style of the Warmups used in Just-in-Time-Teaching (JiTT) \cite{jitt,cm01}.  In either case, students are prompted with a final question asking if they have any questions on the material.  Student responses to this final question form the basis for classroom discussions.

\subsubsection{During Lecture}
\begin{enumerate}[label={\bf \alph*.}]
\item {\bf Mini-lectures}
\end{enumerate}

Since the primary content distribution has occurred pre-lecture, lecture time focuses on mini-lectures addressing specific student questions, practicing the application of concepts and calculations introduced in pre-lecture, and providing student feedback.  Short mini-lectures, in the style introduced in Peer Instruction \cite{pimazur}, focus on the issues brought up by student in their pre-lecture activities.  Rather than using the time to introduce the material, the mini-lecture serves as the vehicle for answering questions.

In principle, students' concerns are available to the lecturer while he/she composes the mini-lecture and students gain ownership of the process when their questions are addressed in the mini-lecture.  In practice, student issues and misconceptions with introductory physics concepts are well known (see, {\it e.g.}, Ref.\ \cite{fiveknight}) and the mini-lecture can be composed before students submit their responses and nevertheless, students gain ownership as their concerns are addressed in the mini-lecture.  In reality, a combination of the two approaches are accessible for the instructor.  Furthermore, the mini-lecture also provides time for lecturers to engage students in the subject matter by presenting applications and live demonstrations.

\begin{enumerate}[resume,label={\bf \alph*.}]
\item {\bf Worksheets}
\end{enumerate}

Most of the remaining time in class is dedicated to student work on activity worksheets.  There are some well-known resources from which to compile material for the worksheets (see, {\it e.g.}, Refs.\ \cite{knightwkbk,mcdermott}).  In addition, we mined our own ``traditional'' lectures -- designed to fill the entire lecture period minus time for a few clicker questions -- for content:  from conceptual questions to problem-solving strategies to applications.  Every important topic in a course is both in a worksheet for students to explore and engage in the material during class time and also assessed in summative assessments.  

In creating these worksheets, we took the most in-depth concepts we wanted students to grapple with and problems we wanted them to solve, identified the skills and conceptual understanding required to do so, and scaffolded the worksheets for students to slowly construct their own understanding of the material.  The process is similar to how one would prepare a well-crafted lecture or write a textbook on the matter, but with a different creative mindset:  instead of using declarative statements to provide the information and the process in an instructor-centered way, one uses questions and prompts in a sort of Socratic dialogue to guide students through the process.  

The worksheets become a de-facto textbook for the course much in the way that {\it Lecture-Tutorials for Introductory Astronomy} \cite{astro101} becomes the de-facto textbook for the Astronomy 101 curriculum promoted by the Center for Astronomy Education \cite{prather04,prb09,prb11}.  We found that students would progress through the worksheets at different rates.  In order to proceed at a pace that most students could keep up with, but to prevent some students from sitting idly after finishing all assigned work, we included additional conceptual and problem solving exercises in each of the topics.  The goal is to encourage students to work throughout the class period and to provide students study resources after class lets out.  In our experience, this is a point of contention that requires instructor intervention to facilitate student buy-in, including a class-wide discussion of this philosophy, a mechanism to distribute answers to worksheet questions (but definitely not solutions to the worksheets), and exam questions mined from worksheet questions not directly discussed in class (especially on the exam).

Logistically, while we were developing worksheets, we would print out the worksheets and distribute them in class.  This was necessary because we would be constantly be developing worksheets before class, just as preparation for a new lecture often includes the time just before the class meeting.  After getting comfortable with the set of worksheets for a specific course, they could be prepared en masse and require students to purchase a copy of the worksheets from the campus bookstore.  This requires students to bring their worksheet book to class, but students quickly adjust to the requirement, especially when they are an integral and valuable aspect of the course.  Additional worksheets or supplemental work can be added to the course by printing and passing them out in class.

Students are instructed to work on the worksheets in groups, to come to a consensus with their group, and to first direct their questions to neighboring groups.  The instructor and the teaching assistant will walk amongst the students in the lecture hall.  Most lecture halls are set up in stadium seating and while not easily adaptable to student group work, students can work in groups with those next to them and immediately above or below them, and instructors can patrol the aisles and occasionally, when necessary, slide toward the middle of a row.

However, when there are two instructors to help hundreds of students, a few strategies are necessary to make the structure work:  (1) the worksheets need to be appropriately scaffolded such that students can make significant strides on their own (especially if they sufficiently prepared before lecture), (2) student groups need to be encouraged to support each other when they have questions, and (3) when the instructors notice a common issue, they should address the class as a whole.  This final strategy is important to both keep the class moving and allow for more difficult work to be done in class.  The intervention can be approached in many ways, including a planned intervention, where the instructors anticipate an upcoming stumbling block, an intervention based on student responses, where a quick visual survey of student worksheets in situ can prompt an instructor intervention, or an intervention based on common student questions, through which general student confusion in particular points on a worksheet can be dealt with as a class.
 
\begin{enumerate}[resume,label={\bf \alph*.}]
\item {\bf Clicker Questions}
\end{enumerate}

To conclude a session of worksheet work, we introduce one or more clicker questions in a think-pair-share style to encourage students to think about and discuss the main point(s) of the worksheet covered and as a mechanism to get students on the same page.  This provides a last chance to revisit the sticky topics covered and to give students an opportunity to ask questions they may have.

The fifty-minute lecture period is easily divided into two cycles of mini lecture-worksheet-clicker questions.  Our goal is to spend at least half of lecture time with students working on worksheets or clicker questions, which accentuates the importance of student work while de-emphasizing the role of the mini-lecture.  Through our experiences and the experiences of others who tried to mimic our techniques, we found that this was an important aspect of student buy-in and the resultant student learning. 
``Less is more'' became our mantra in preparing the mini-lecture, to cover material that truly couldn't be done through a worksheet, including demonstrations and other neat applications of the material and lecture aimed to specifically address important misconceptions and difficulties students face in the material.  We found that students would prepare for class, knowing that the basics wouldn't be covered in the mini-lecture and that they would be lost without some amount of preparation.

\begin{enumerate}[resume,label={\bf \alph*.}]
\item {\bf Wrap Up}
\end{enumerate}

The last five minutes of lecture might not sound terribly important, but having a wrap-up activity of some sort during this time is critical.  This time can be used in many ways, but our general method is to have students write and submit a ``minute paper'':  students write their answers and explanations to a specific worksheet or lecture question on an index card for submission and feedback.  These cards are then graded on an effort-based scale by the next lecture.

This activity is valuable since (a) it forces students to \emph{create} content for feedback, (b) it recognizes that students need to take the worksheet questions seriously, (c) it allows the instructor to gauge class-wide understanding on specific parts of the worksheet, and (d) it guides the instructor on how to plan for the next class period.  We have found that this activity further serves to hold students responsible for the material, to encourage them to keep up with the material, and to serve as a starting point for a number of out-of-class discussions.

Index cards are useful because they are easily passed out and collected and they are easily sorted into categories to assess student responses.  One can quickly flip through a class's notecards to discern major student issues to address in the next class period, and they can be easily sorted if one is interested in making these responses a part of the course grade or providing limited feedback to students on their responses.

\begin{table*}[t]
\begin{tabular}{| p{3.5in} | p{3.5in} |}
\hline
{\bf Traditional} & {\bf Flipped} \\
\hline \hline
Suggested reading from textbook & Suggested reading from textbook \\
 & Instructor created outlines of material \\
\hline
 & JiTT-style open-ended reading quizzes \\
 & including ``what questions do you have for me?'' \\
\hline
2-3 clicker questions during lecture & 2-3 clicker questions during lecture \\
Think-Pair-Share associated with clicker questions & Think-Pair-Share associated with clicker questions \\
\hline
{\it Instructor-driven lecture} & {\it Student-centered activity worksheets} \\
$\gtrsim 80\%$ of class time in instructor-driven lecture & $\lesssim 50\%$ of class time in instructor-driven lecture \\
\hline
Suggested (ungraded) homework from textbook & Suggested (ungraded) homework from textbook \\
\hline
Bi-weekly multiple choice quizzes & Bi-weekly multiple choice quizzes\\
Multiple choice final exam & Multiple choice final exam \\
\hline
\end{tabular}
\caption{A comparison of the ``traditional'' and ``flipped'' classrooms used in this study.  The ``traditional'' classes utilize some active-engagement, student-centered methods, but not to overall level as seen in the ``flipped'' classes.  \label{tab:details}}
\end{table*}

\subsubsection{After Lecture}

Flipping the classroom creates an environment that promotes student engagement in the material under their instructor's guidance. However, in order for students to improve their understanding of the material, they will need to devote time and effort outside of the classroom hour.  In all of our flipped courses (just as in our traditional unflipped courses), students were offered a set of additional homework problems that were not graded (and were thus, treated by many students as optional).  This choice was rooted in the resources available, but online homework could be employed as well, as long as those problems are chosen to work in concert with the worksheets and learning goals of the class.  Too often, we found that online homework from the textbook's online system would diverge from our specific learning goals and methods, so we did not use them.  The connection between the homework and the learning goals is important so that this work is seen as valuable, and not merely busywork.

Students also will find a need to get help more often as their ``course textbook,'' {\it i.e.}, the worksheets, require continued discussion.  Discussion boards, tutoring times, and help sessions are all useful modalities to supplement and kindle that discussion.  We also found more meaningful interactions in office hours than we've previously experienced as students come in with more directed and pertinent questions.  In essence, the worksheets ``prime the pump'' for relevant and meaningful student inquiry into the material.

\section{Analysis}

This section compares student results in both traditional lectures and classes using our flipped methodology taught by the same instructors.  Note that while the ``traditional'' classes were taught using some active-engagement methods (including 2-3 clicker questions presented with think-pair-share methods), most of the lecture time ($\gtrsim 80\%$) was dedicated to instructor-centered lecture.  Table \ref{tab:details} outlines a comparison of similarities and differences between the two.  To adjudicate the efficacy of the flipped classroom methodology, we analyzed results of multiple choice questions on quizzes and final exams administered in the flipped classrooms we taught and in the traditional classrooms we taught before beginning to employ the flipped pedagogy.  All plans for analysis were made {\it after} the conclusion of all courses that have been analyzed.

For each multiple choice question, we want to find the fraction of students $p$ that would correctly answer the question in a typical class, taught by the particular instructor using a given classroom methodology.  However, because we only have data to analyze after the fact, we analyzed the results to estimate this fraction and the uncertainty on this fraction, $\sigma_p$.  We used the fraction of the class with correct responses to the question to estimate $p$ and treated each multiple choice question as a random binomial variable with probability $p$, so we could estimate the uncertainty as
\begin{equation}
\sigma_p = \sqrt{\frac{p(1-p)}{N}} ,
\end{equation}
where $N$ is the number of students in the class.  The class sizes are on the order of one to three hundred, so these are likely appropriate estimates of $p$ and $\sigma_p$. 

The results that follow involve students in the three-quarter introductory physics sequence for students in the life- and health-sciences and are distinguished into the three quarters:  Newtonian mechanics (Mechanics), electricity and magnetism (\eandm), and oscillations, waves and modern physics (Waves/Modern).  Throughout this section, the figures will be coded with red circles representing results from Mechanics, blue triangles representing results from \eandm, and green stars representing results from Waves/Modern.

\subsection{Identical Questions}

We first compare identical multiple choice questions that were given in both the traditional and flipped classes during a quiz or the final exam.  These were questions that were given by the same instructor in the same course, but in different academic terms.  By {\it identical}, we mean that the questions differ in superficial ways including differences only in the numbers used in a calculation or cardinal directions introduced in a conceptual question.  For this comparison, we introduced as stringent a cut on the data as possible to perform as much of an apples-to-apples comparison between the two classes.  In this section, this hard cut meant that we did not analyze pairs of questions that were similar in set up (including those with the same physical setup but students were asked to solve for a different variable), contained additional information (including adding unnecessary information or a diagram), or were isomorphic in physics content or mathematical structure.

\begin{figure}[t]
\includegraphics[width=3 in]{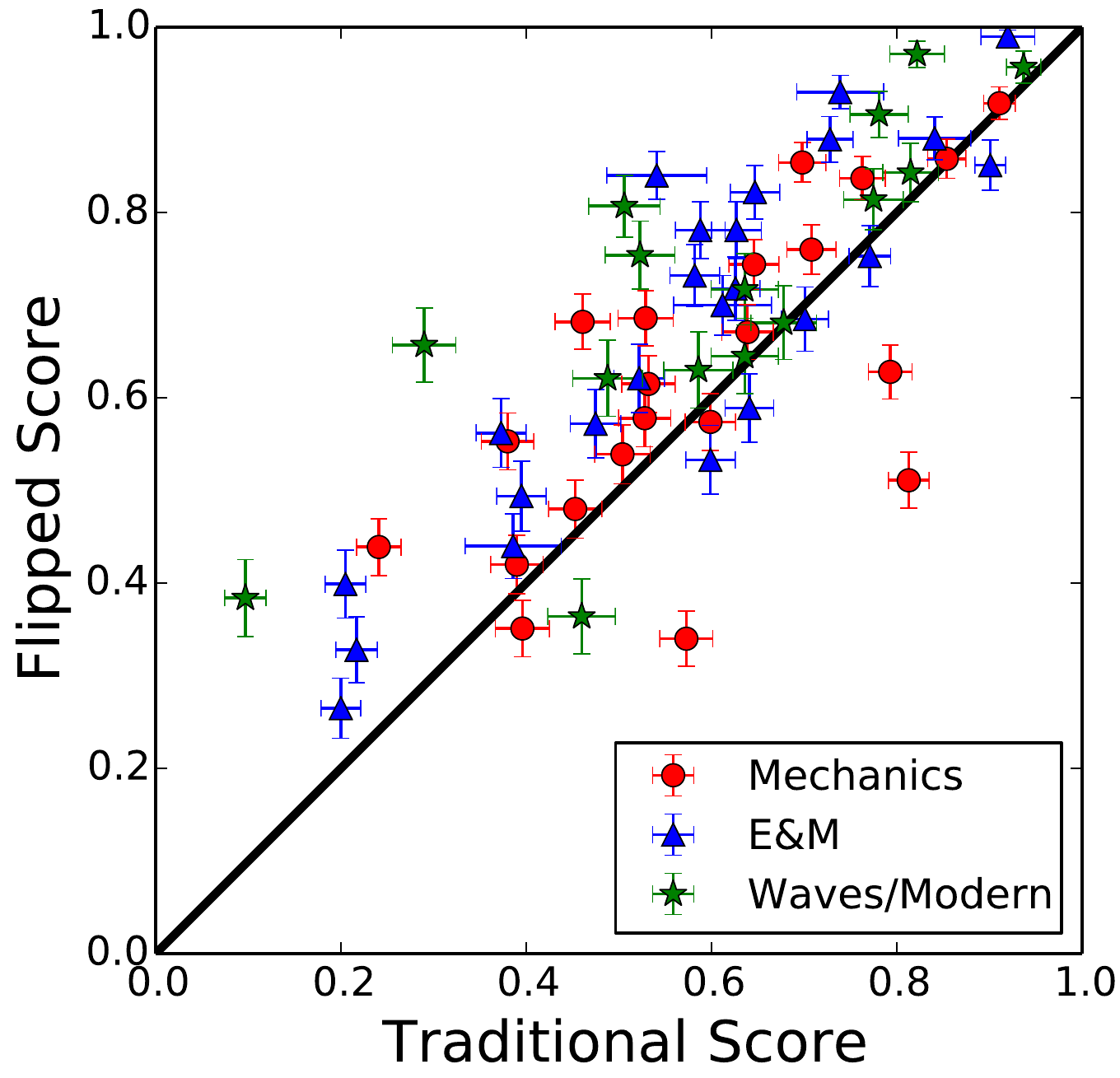}
\caption{A comparison of identical questions administered as a part of a quiz or final exam in both a traditional class and a flipped class.  The solid, $45^\circ$ line indicates equal student performance in flipped and traditional classroom settings.  Points above the line indicate better performance in flipped classrooms, while those below the line represent poorer performance in flipped classrooms on identical question.  }
\label{fig:compq}
\end{figure}

Figure \ref{fig:compq} shows the results of this comparison.  The flipped and traditional scores are the fraction of students who correctly answered each identical question.  The $45^\circ$ line represents equal student performance in flipped and traditional classrooms on the identical questions.  Points above the line represent questions where students performed better in the flipped than the traditional class.  In all, the figure represents 60 identical questions from across seven flipped sections and ten traditional sections.  The hard cut on the data has reduced the number of questions analyzed in this comparison by at least an order of magnitude from the quiz and final questions administered in the flipped courses.

A few results can be visually deduced from Figure \ref{fig:compq}:  (a) Four of the sixty points ($\sim 7\%$) lie significantly below the line, meaning that for almost every question, student in the flipped classes performed at or above the level of students in traditional classes.  (b) While many of the questions are consistent with the $45^\circ$ line, over half of the points are significantly above the line, meaning that on these questions, students in flipped classes performed significantly better than those in traditional classes.  (c) The points most significantly below the line are from Mechanics.  We will speculate about the reason for this last point in the Discussion section, but the first two points indicate that in this most direct comparison, students in our flipped classes overall performed significantly better than those in our traditional classes.

The identical questions comparison provided a means of comparing classes between academic terms since we were able to control for the question, the instructor and the course without having to independently assess the difficulty of questions.  While we did not perform the analysis shown in Figure \ref{fig:compq} until long after the course was completed, it became immediately obvious that students in the flipped classes were outperforming the students in our previous traditional classes.  This resulted in the need to introduce both similar questions to those we administered in the past, but also more difficult questions that were either richer conceptually or asked students to perform more in-depth calculations.

\subsection{Gender Comparisons}

\begin{figure*}[t]
\includegraphics[width=6 in]{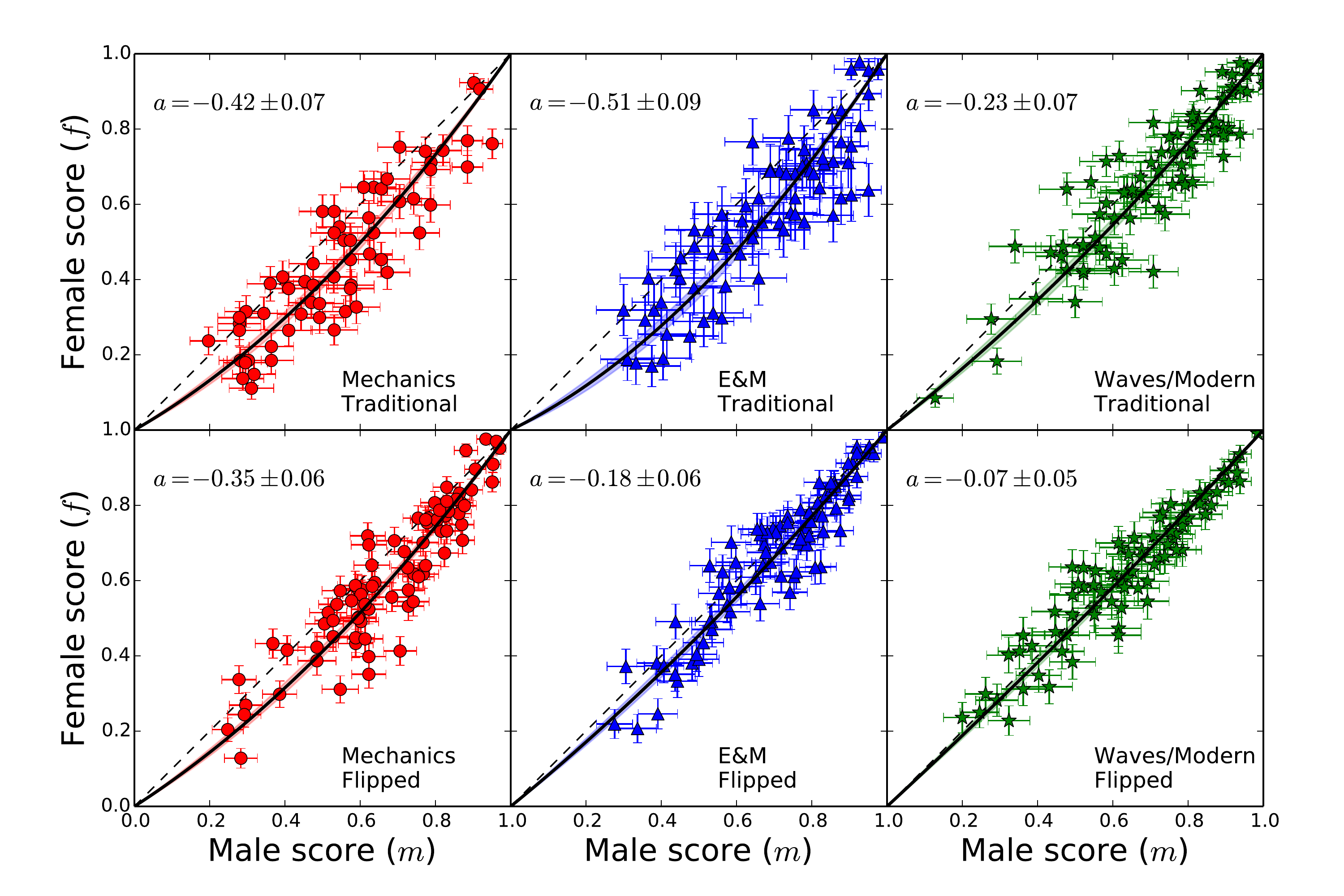}
\caption{A comparison of female and male scores on each question given in a course during a quiz or the final throughout a single academic term.  Each of the six subplots contain the results of every question given in a course.  Vertical comparisons are the same course, with the same instructor, but with a different instructional pedagogy and during a different quarter.  The overlaid model is the best fit to the gender gap model, Eq.\ (\ref{eq:gendergapmodel}).  The dashed $45^\circ$ line is equal scores for male and female students, overall representing no gender gap.}
\label{fig:gendercomp}
\end{figure*}

Figure \ref{fig:gendercomp} represents female scores ($f$) versus male scores ($m$) on each of the $\sim 80$ questions in the five quizzes and the final exam given in an individual course in a given academic term (with the same instructor and same instructional methodology).  In each of the six subplots, every point represents a different question.  Vertically separated plots compare the same course ({\it e.g.}, Mechanics, \eandm, Waves/Modern) with the same instructor, but different methodology (traditional above vs. flipped below) in different academic terms. 

We noticed in these plots that in each course there is evidence of a gender gap:  there are significantly more questions in each course where male students answered the questions correctly at a higher fraction than female students than vice versa.  While this gap has been observed in the literature, even when active learning pedagogies are used in the classroom \cite{lcm06,pfk07,kpf09,mms13,hen17,kms18}, it is certainly not an ideal outcome of any teaching methodology.  When comparing the traditional classes to the flipped classes, we noticed that in the flipped courses, the data points tended to group closer to the $45^\circ$ line, indicating a lesser discrepancy between male and female scores in the course.

To analyze this trend we introduced a mathematical model:
\begin{equation}
f - m = a m (1 - m) ,
\label{eq:gendergapmodel}
\end{equation}
which is a one-parameter model to estimate the gender gap between the overall performance of female to male scores on all questions in a course.  $a = 0$ represents no gender gap with equal performance between the genders, while $a < 0$ represents a gender gap between female and male performance in the course, where more negative values of $a$ represent a larger gender gap.  
We call the parameter $a$ the gender gap parameter because the more negative it is, the worse it is.  (In principle, $a$ could be positive, indicating a gender gap in the other direction, which could be equally problematic.  However, in practice we have not seen this case, nor does there appear to be literature in physics education indicating this opposite gender gap.)

This model seems to mimic the data well, especially at the end points where $f \approx m \approx 0$ and $f \approx m \approx 1$ and in the middle of the plot where more negative values of $a$ represent a greater dip of the data below the $45^\circ$ line.  The model seems appropriate for the data sets in this study, but would be ill-suited to the most egregious gender gaps where these limiting cases break down:  when a gender largely gets a question correct ($m ~{\rm or}~ f \approx 1$) while the other does not, or when a gender largely gets a question wrong ($m ~{\rm or}~ f \approx 0$) while the other does not.

To estimate the gender gap parameter for each course we performed a $\chi^2$-minimization of the course data with respect to the one-parameter model, Eq.\ (\ref{eq:gendergapmodel}) \cite{numrec}.  In performing this minimization, we treated our uncertainties as if they were approximately Gaussian.  In addition, we attempted to ascertain uncertainties on this parameter, recognizing not only the error bars on the individual questions, but different instances of the same course with the same instructor and same methodology may have a different ensemble of questions.  To do this, we first assume that the questions analyzed for each course are representative of the instructor's exam questions in representative proportions of conceptual/calculations and of relative difficulties (in order to perform the analysis done in this subsection, we have already implicitly made this assumption). We performed a bootstrap \cite{numrec} on the questions, re-analyzing each stochastic instantiation of the data, using the $\chi^2$-likelihood of the parameter given this data, $\mathcal{P} (a) \propto e^{-\chi^2/2}$.  After a large number of stochastic instantiations of the data, the likelihood of $a$ given the data is approximately Gaussian, and we use the mean and standard deviation of this Gaussian to estimate the value of the gender gap parameter and its uncertainty for a course.  The result is a greater uncertainty on $a$ than would have been inferred from a $\chi^2$-minimization alone, which is reasonable since we are only able to assess a course once.

\begin{figure}[t]
\includegraphics[width=3 in]{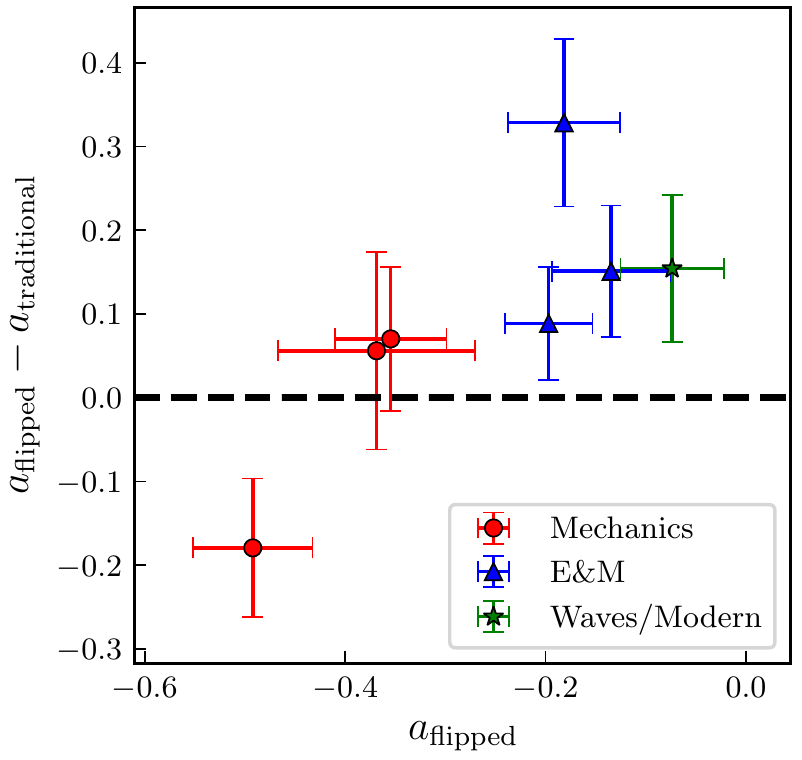}
\caption{Comparison of the gender gap parameter for traditional, $a_{\rm traditional}$, and flipped classes, $a_{\rm flipped}$, with the same instructor and same course, but during a different academic term.  The ordinate is the difference in the parameter between the flipped and traditional classes.  The horizontal dashed line is no difference in the gender gap, while points above the line represent a reduction in the gender gap and points below the line represent a worsening of the gap.}
\label{fig:acomp}
\end{figure}

Qualitatively, the gender gap parameter $a$ is generally closer to zero for each of the flipped classroom fits as compared to the traditional fits, indicating that our flipped classroom methodology significantly reduced the gender gap.  Figure \ref{fig:acomp} embodies our attempt to quantify this trend by comparing the difference in the gender gap parameter between flipped and traditional courses taught by the same instructor over different quarters. 
In this figure, the difference in the gender gap parameter in the flipped course and the traditional course is plotted versus the parameter for the flipped course alone.  The ordinate of the graph is the {\it reduction} of the gender gap in the flipped classroom, so positive values (above the horizontal black dashed line) represent improvement in flipped classes relative to a traditional class controlling for the instructor and the course.  This figure summarizes the results for the gender gap parameter from Figure \ref{fig:gendercomp} along with four other flipped courses.

There are a few results that can be immediately observed from Figure \ref{fig:acomp}.  (a) Most courses are above the horizontal line, representing a reduction in the gender gap in the flipped course.  (b) It is notable that the exception is a Mechanics course (again, see the Discussion section for a discussion of mechanics and the methodology described in this work).  (c) The gender gap parameter is still negative for all flipped courses analyzed, indicating that although there seems to be a general trend of reducing the gender gap, the gap remains.  (d) For the same course, the gender gap parameter for flipped courses $a_{\rm flipped}$ seems to be consistent between the instructors analyzed in this work.  This appears to be a validation of the assumptions made above to estimate the value of the parameter and its uncertainty.

\section{Discussion}

%
A general take away from the data in the previous section is that, with a few exceptions, students in the flipped courses performed better on quizzes and the final exam than students in the traditional courses, and we infer this means that students have learned more in these courses as well.  It should be noted that the exceptions to the general take-away involve the mechanics courses.  In this section we will first speculate why the analysis of the mechanics courses have provided some evidence contrary to this general take away, 
then share some of the pitfalls that we have encountered along the way.

\subsection{Why Mechanics?}

In Figure \ref{fig:compq}, the identical questions where the flipped class scores were most significantly less than those in the traditional class were from mechanics courses.  These three questions most below the $45^\circ$ line are both conceptual and calculation based and from different sections in the curriculum.  However, it should be noted that while these three points are the most eye-catching in the figure, they represent roughly $20\%$ of the total mechanics questions on the graph.  So, in the flipped classroom, most of the identical questions lead to either significantly positive gains or no significant gains in student scores.

In Figure \ref{fig:acomp}, the \eandm{} and Waves/Modern courses show significant improvement in the reduction of the gender gap, while the mechanics course results are generally consistent with no  overall improvement, and one course reported a worsening of the gender gap.  Taken as a whole, it should be noted that there are positive gains made in the flipped classroom in the mechanics courses, and at worst, the data are consistent with no overall harm with the introduction of the flipped classroom to our mechanics courses.

So, a natural question is why is mechanics an outlier?  Although we do not have further data to analyze this question, we can speculate the cause:
\begin{itemize}
\item It is the first course in a three-quarter introductory sequence, but most of the students are not incoming first year students, rather they are life- and health-sciences major students taking physics in their third or fourth years.
\item Some students are only required to take this first quarter to satisfy their major's requirements.
\item Many students have never taken a physics course before; while many students have also taken Advanced Placement or International Baccalaureate (AP or IB) physics before taking the class.  This means that there are significant numbers of students who have not been introduced to the fundamental mathematical and physical skills needed in the mechanics class, and also significant numbers of students who have from their previous (high school) coursework.
\item Those who have had physics courses in high school come from programs that have greatly emphasized mechanics with brief excursions into electromagnetism and rarely cover waves and modern physics.  
\end{itemize}

Certainly, one cannot discount the effect of the first two points.  The culture shock of students, often students advanced in their college career, taking their first college physics course can be pointed to as a contributing factor in students' struggle in mechanics.  Moreover, for students where mechanics is also their final college physics course, they have been disincentivized from building a strong base of fundamentals that is built in mechanics since they will not take the rest of the sequence and by their major departments who have devalued the physics sequence by only requiring the first term in the sequence.  However, it would be defeatist to accept those as insurmountable obstacles that cannot be overcome by enhanced focus on student buy-in and improved methodologies.  

The final two points are related to student pre-conceptions upon entering the course.  Even without previous physics course, students' interaction with Nature throughout their lifetime before entering the classroom means that they enter this first college physics course with the most pre-conceptions of the three courses in the sequence.  In mechanics, they are the least {\it tabula rasa}.  Among these pre-conceptions are many student misconceptions, both subtle and egregious.  The goal of our flipped classroom pedagogy (and one could include most of the physics education active learning enterprise) is to engage these pre-conceptions and through pre-class preparation, scaffolded worksheets and class-wide interventions and formative assessments.  Without this engagement, students' misconceptions will remain \cite{learn}, inhibiting student learning in mechanics and reducing student scores on summative assessments.

Certainly students are not a tabula rasa when they enter the \eandm{} and Waves/Modern classes, but these pre-conceptions are often less-solidly ingrained in their understanding of the world and there are typically fewer of these pre-conceptions.  The breadth and depth of these pre-conceptions in the mechanics class does not alone make it more difficult to address and engage the misconceptions, but the significant variance in students' previous physics preparation for the course exacerbates the problem.

We have found that the mechanics course has the greatest variance in the rate at which students work through activity worksheets.  The greater number of pre-conceptions along with the variance in rates means that some student groups may have breezed through multiple pre-conceptions in the time span when other groups haven't yet resolved their first.  While the goal of a well-scaffolded worksheet is to work students through these pre-conceptions to resolve any misconceptions the students may bring into the class, they rely on instructor intervention to keep students on track and further engage and address these misconceptions.  However, the instructor intervention is effective if it occurs after students have had the opportunity to engage in the material.  This leads to unsavory options including waiting until the last (or, perhaps $n$th-percentile) student group completes their work on each concept addressed while the faster groups are either waiting around or re-practicing the same material, or having the fastest groups move ahead and rely on the scaffolding of the worksheet to keep them on track allowing for later interventions on the material.  We have chosen the latter in this study.  

It remains to be seen what effect the student population has on the discrepancy between mechanics and the other courses in the sequence, what effect the broad range of previous student experiences in physics classes and the resulting range of rates of student work on the worksheets has on this discrepancy or whether other factors contribute as well.  We continue to look at the role of student buy-in in addressing the first issue and we will continue to experiment with the methodology to address the second.

%

%

\subsection{Lessons Learned}

In creating this flipped classroom methodology for large lecture courses with minimal departmental/university resources, we have learned many lessons the hard way:  (1) worksheet solutions should \emph{never} be made, (2) student buy-in is incredibly important, but also requires constant attention, (3) worksheets should permeate through all aspects of the course, and (4) creating worksheets is a time-consuming yet valuable process that gets easier with experience.

\begin{enumerate}[label={\bf \arabic*.}]
\item {\bf No worksheet solutions}
\end{enumerate}

The activity worksheets serve as the de-facto course textbook.  Therefore, students will feel that it is a reasonable request for worked solutions to the worksheets.  In our experience, creating worked solutions completely short-circuits the learning process that the flipped classroom works so hard to achieve.  The allure of the worked solutions will break many students' will to persevere through the frustrating process of learning, especially as quizzes provide a strong incentive to ``learn the correct answer,'' rather than to slog through the process of breaking through misconceptions to slowly (and often frustratingly) reach the correct answer.  In addition, scaffolded worksheets are time-consuming to make and creating worked solutions means that ``official'' solutions will forever exist for your worksheets, short-circuiting the learning process even more in future academic terms.

We have strongly opposed acquiescing to these student requests and creating worked solutions for mass distribution through the course homepage.  In our experience, students are most passionate about this request, and this topic needs to be broached with students in class (or it will show up on your teaching evaluations in a negative way).  In this request, some students are being metacognitive in their learning process, and this metacognition needs to be met with respect and dialog about why there cannot and will not be worked solutions.  Suggestions for other mechanisms for students to provide feedback for their worksheet work should be both offered and taken.

As a ``compromise,'' students need to be offered some means of knowing that they are on the right track as they work through their worksheets on their own.  It is unreasonable to expect that the hundreds of students in the class will be able to make it to the professors office hours for the specific reason of checking their worksheets.  One useful technique is to offer the correct answers to select worksheet questions, in the same way that textbooks have answers in the back of the book.  This allows students to complete the worksheet and compare their answers, creating a mechanism for them to know that they need help.  Finally, the course's teaching assistant(s) need to know this prohibition against creating worksheet solutions, lest they create them with the best intentions of helping students while actually hurting their learning processes.

\begin{enumerate}[resume,label={\bf \arabic*.}]
\item {\bf Student Buy-in}
\end{enumerate}

It is no secret amongst those who teach in an active classroom that student buy-in is an essential part of creating the classroom culture needed for active learning to be successful.  In a large introductory lecture hall, especially at a research-focused university, students have grown accustomed to being anonymous as they passively attempt to learn from a lecturing instructor.  It is no surprise that these very students are often resistant to new modes of teaching and learning, especially as most introductory science courses do not employ active techniques that take up most of the class period.  

Working toward student buy-in requires constant work and attention on the instructor's part.  We do not cover physics content on the first day of classes, rather spending time detailing the way the flipped classroom works and most importantly why we do the things we do.  In addition, more time is required throughout the course to emphasize the process, why we do the process, and ways students can better interact with the process.  An early first quiz can be an effective means to engage students in the learning process, by pointing out connections between the exam and the worksheets.  Students must see the worksheets as being an integral part of their success on the quizzes and exams or they will not buy in to replacing lectures with student-driven worksheets.

\newpage
\begin{enumerate}[resume,label={\bf \arabic*.}]
\item {\bf Worksheets should permeate the course}
\end{enumerate}

An important step in building student buy-in is to make worksheets permeate all aspects of the course.  Pre-lecture materials aim to prepare students for the work they will do on their worksheets.  The mini-lecture in class aims to address common questions students will have (and have exhibited through their pre-lecture assignments) regarding the worksheet material or aims to demonstrate the concepts students have worked on in their worksheets.  Clickers, minute papers, and any other in-class formative assessments should reflect on the worksheet material.  And most importantly, homework and quizzes should reflect worksheet material.

The instructor indicates to the students the importance of the worksheets not only by stating that it is so, but also by structuring the course around these worksheets.  Effectively, this means that the graded elements (homework and quizzes) need to reflect students' work on the worksheets.  A subtle, yet powerful way that instructors inadvertently can discount the value of the worksheets is by reducing the amount of class time spent working on the worksheets.  With honest and good intentions, the instructor's natural instinct is to explain {\it more} when a difficult subject is upcoming and to attempt to take more control of the class structure when students face issues.  However, reducing the time spent on worksheets reduces the worksheet's importance to the students, and further convinces the students that they {\it need} to be lectured to in order to learn the material (reinforcing a prevalent pre-existing student attitude).  Losing a few minutes here or there to put students on the same page probably isn't harmful, but if one isn't careful, it can spiral to the point where worksheets take a continually decreasing fraction of class time and can lead to a vicious spiral where students demand more and more lecturing, reducing the time for worksheets and reducing the importance of the worksheets in students' eyes.

Worksheets should be done every day and for a significant fraction of the class period, without exception.  One study \cite{lsci} found that a threshold fraction of the class time needed to be spent on student-centered activities was a necessary (but not sufficient) condition for significantly larger student conceptual gains in introductory astronomy courses.  While there are some differences between the introductory physics and introductory astronomy curricula, the basic conclusions are likely generally true, if not the specifics.

\begin{enumerate}[resume,label={\bf \arabic*.}]
\item {\bf Creating Worksheets}
\end{enumerate}

\begin{quote}
If a topic is important, it needs to be in a worksheet, and thus must be assessed as a part of student grades. 
\end{quote}

Worksheets need to address the learning goals of the instructor, scaffolded in a way that the topics covered are accessible to students without heavy facilitation, and be significant enough to be done every day and for a significant fraction of the class.  It should come as no surprise that creating these worksheets is time consuming.  We found that preparing the worksheets for a course that we've already taught took at least the same amount of time as preparing for an entirely new course.  

Converting an existing lecture to a worksheet involves converting: (1) declarative statements made by a lecturer to questions on a worksheet, (2) algebraic steps worked out on the board or in a powerpoint slide to be guided steps on a worksheet, and (3) questions asked to the class to either be a part of a guided Socratic questioning in a worksheet or an instructor-led intervention, such as a clicker question.  Moreover, existing published resources ({\it e.g.}, Refs.\ \cite{knightwkbk,mcdermott}) can be converted to worksheet questions with minimal difficulty, and can be modified to suit the instructor's goals by increasing student guidance in a worksheet or focusing students toward the primary lesson of the worksheet.  It should be noted that even an average worksheet that focuses on the learning goals and requires students to engage in the material will be more beneficial to student learning than a full period focused on instructor-centered lecture.

%




\section{Concluding Thoughts}

%

In conclusion, the data suggest that the large lecture classroom can be flipped to replace instructor-centered lecture with student-driven activities while still respecting constraints on physics courses (and constraints on the time of physics faculty) at large universities. 
Even under these constraints, a flipped classroom implementation still diminishes the widely-observed gender gap in introductory physics courses, as seen across a wide array of active-learning methods \cite{freeman14}.

Ref.\ \cite{lcm06} lists a number of strategies to reduce the gender gap.  Our flipped class implementation uses and addresses a number of these strategies, especially the creation of an interactive environment that enhances cooperation and communication between students and with the instructor, that alternates between group discussion and structured teaching, with activities that decrease competitiveness and utilizes diverse and frequent assessment and feedback.
However, these strategies may be less effective without {\it also} addressing the social dynamics of the classroom.  One study in biology found that students tend to exhibit a gendered bias when anonymously nominating knowledgable students in their class \cite{gebwcg}.  The authors of the study suggest that their result may be a manifestation of implicit bias in STEM majors \cite{nbg02,ns11}.  Furthermore, one of the specific suggestions they offer to reducing this bias is through utilizing student-centered activities that focus on small group work rather than whole class instruction.  This is consonant with the fundamental tenets of our methodology.

Yet, we still observe a gender gap in our classroom, albeit significantly reduced in some cases.  As we look forward, we ask whether this is a result of our implementation of the methodology, an indicator of structural changes that could be made to the methodology, or perhaps may be related to the social culture of STEM disciplines at any given institution.  One possibility may be that the social dynamics observed on the class-wide scale ({\it e.g.}, Ref.\ \cite{gebwcg}) may be going on to a lesser extent in small groups.  One strategy to approach this issue could be to discuss with students the cognitive and sociological effects that lead to implicit biases to encourage them to be self-aware in their classroom discussions.  While the first day of class is an appropriate time to begin this discussion, it will never-the-less be important to follow up throughout the term requiring either time in class or a brief reading and/or reflection assignment mid-term.

 
The analysis performed in this work can be applied to other traditionally under-represented groups.  Our dataset was limited to gender; because the data was collected without this analysis in mind, our options to test other demographic parameters were limited.  While testing other demographic factors, it should be noted that the analysis will suffer from having small numbers of a tested group.  Perhaps the specific institution, in the particular classes where this analysis was performed may not be appropriate for further analyses because of the specific demographics of the classes, but other institutions may present better opportunities to use the same analysis for race/ethnicity, sexuality, age, etc.


Personally, we have found that the flipped classroom methodology as described in this manuscript has so transformed every aspect of our courses with a powerful combination of student learning gains, improved student engagement, and energy throughout every aspect of the learning process, that we cannot in good faith foresee returning to a traditional lecture structure.

\acknowledgements

We would like to thank I.\ Schanning for his work in developing the student activity worksheets and K.\ Tanner for useful discussions.

\bibliography{flip}

\begin{thebibliography}{35}
\expandafter\ifx\csname natexlab\endcsname\relax\def\natexlab#1{#1}\fi
\expandafter\ifx\csname bibnamefont\endcsname\relax
  \def\bibnamefont#1{#1}\fi
\expandafter\ifx\csname bibfnamefont\endcsname\relax
  \def\bibfnamefont#1{#1}\fi
\expandafter\ifx\csname citenamefont\endcsname\relax
  \def\citenamefont#1{#1}\fi
\expandafter\ifx\csname url\endcsname\relax
  \def\url#1{\texttt{#1}}\fi
\expandafter\ifx\csname urlprefix\endcsname\relax\def\urlprefix{URL }\fi
\providecommand{\bibinfo}[2]{#2}
\providecommand{\eprint}[2][]{\url{#2}}

\bibitem[{\citenamefont{Halloun and Hestenes}(1985)}]{hh85}
\bibinfo{author}{\bibfnamefont{I.~A.} \bibnamefont{Halloun}} \bibnamefont{and}
  \bibinfo{author}{\bibfnamefont{D.}~\bibnamefont{Hestenes}},
  \bibinfo{journal}{Am.\ J.\ Phys.} \textbf{\bibinfo{volume}{53}},
  \bibinfo{pages}{1043} (\bibinfo{year}{1985}).

\bibitem[{\citenamefont{McDermott}(1991)}]{mcd91}
\bibinfo{author}{\bibfnamefont{L.~C.} \bibnamefont{McDermott}},
  \bibinfo{journal}{Am.\ J.\ Phys.} \textbf{\bibinfo{volume}{59}},
  \bibinfo{pages}{301} (\bibinfo{year}{1991}).

\bibitem[{\citenamefont{Hake}(1998)}]{hake98}
\bibinfo{author}{\bibfnamefont{R.~R.} \bibnamefont{Hake}},
  \bibinfo{journal}{Am.\ J.\ Phys.} \textbf{\bibinfo{volume}{66}},
  \bibinfo{pages}{64} (\bibinfo{year}{1998}).

\bibitem[{\citenamefont{Deslauriers et~al.}(2011)\citenamefont{Deslauriers,
  Schelew, and Wieman}}]{dsw11}
\bibinfo{author}{\bibfnamefont{L.}~\bibnamefont{Deslauriers}},
  \bibinfo{author}{\bibfnamefont{E.}~\bibnamefont{Schelew}}, \bibnamefont{and}
  \bibinfo{author}{\bibfnamefont{C.}~\bibnamefont{Wieman}},
  \bibinfo{journal}{Science} \textbf{\bibinfo{volume}{332}},
  \bibinfo{pages}{862} (\bibinfo{year}{2011}).

\bibitem[{\citenamefont{Docktor and Mestre}(2014)}]{synthesis}
\bibinfo{author}{\bibfnamefont{J.~L.} \bibnamefont{Docktor}} \bibnamefont{and}
  \bibinfo{author}{\bibfnamefont{J.~P.} \bibnamefont{Mestre}},
  \bibinfo{journal}{Phys. Rev. ST Phys. Educ. Res.}
  \textbf{\bibinfo{volume}{10}}, \bibinfo{pages}{020119}
  (\bibinfo{year}{2014}).

\bibitem[{\citenamefont{Freeman et~al.}(2014)}]{freeman14}
\bibinfo{author}{\bibfnamefont{S.}~\bibnamefont{Freeman}} \bibnamefont{et~al.},
  \bibinfo{journal}{Proc.\ Nat.\ Acad.\ Sci.\ USA}
  \textbf{\bibinfo{volume}{111}}, \bibinfo{pages}{8410} (\bibinfo{year}{2014}).

\bibitem[{\citenamefont{Lorenzo et~al.}(2006)\citenamefont{Lorenzo, Crouch, and
  Mazur}}]{lcm06}
\bibinfo{author}{\bibfnamefont{M.}~\bibnamefont{Lorenzo}},
  \bibinfo{author}{\bibfnamefont{C.~H.} \bibnamefont{Crouch}},
  \bibnamefont{and} \bibinfo{author}{\bibfnamefont{E.}~\bibnamefont{Mazur}},
  \bibinfo{journal}{Am.\ J.\ Phys.} \textbf{\bibinfo{volume}{74}},
  \bibinfo{pages}{118} (\bibinfo{year}{2006}).

\bibitem[{\citenamefont{Pollock et~al.}(2007)\citenamefont{Pollock,
  Finkelstein, and Kost}}]{pfk07}
\bibinfo{author}{\bibfnamefont{S.~J.} \bibnamefont{Pollock}},
  \bibinfo{author}{\bibfnamefont{N.~D.} \bibnamefont{Finkelstein}},
  \bibnamefont{and} \bibinfo{author}{\bibfnamefont{L.~E.} \bibnamefont{Kost}},
  \bibinfo{journal}{Phys.\ Rev.\ ST Phys.\ Educ.\ Res.}
  \textbf{\bibinfo{volume}{3}}, \bibinfo{pages}{010107} (\bibinfo{year}{2007}).

\bibitem[{\citenamefont{Kost et~al.}(2009)\citenamefont{Kost, Pollock, and
  Finkelstein}}]{kpf09}
\bibinfo{author}{\bibfnamefont{L.~E.} \bibnamefont{Kost}},
  \bibinfo{author}{\bibfnamefont{S.~J.} \bibnamefont{Pollock}},
  \bibnamefont{and} \bibinfo{author}{\bibfnamefont{N.~D.}
  \bibnamefont{Finkelstein}}, \bibinfo{journal}{Phys.\ Rev.\ ST Phys.\ Educ.\
  Res.} \textbf{\bibinfo{volume}{5}}, \bibinfo{pages}{010101}
  (\bibinfo{year}{2009}).

\bibitem[{\citenamefont{Madsen et~al.}(2013)\citenamefont{Madsen, McKagan, and
  Sayre}}]{mms13}
\bibinfo{author}{\bibfnamefont{A.}~\bibnamefont{Madsen}},
  \bibinfo{author}{\bibfnamefont{S.~B.} \bibnamefont{McKagan}},
  \bibnamefont{and} \bibinfo{author}{\bibfnamefont{E.~C.} \bibnamefont{Sayre}},
  \bibinfo{journal}{Phys.\ Rev.\ ST Phys.\ Educ.\ Res.}
  \textbf{\bibinfo{volume}{9}}, \bibinfo{pages}{020121} (\bibinfo{year}{2013}).

\bibitem[{\citenamefont{Henderson et~al.}(2017)\citenamefont{Henderson,
  Stewart, Stewart, Michaluk, and Traxler}}]{hen17}
\bibinfo{author}{\bibfnamefont{R.}~\bibnamefont{Henderson}},
  \bibinfo{author}{\bibfnamefont{G.}~\bibnamefont{Stewart}},
  \bibinfo{author}{\bibfnamefont{J.}~\bibnamefont{Stewart}},
  \bibinfo{author}{\bibfnamefont{L.}~\bibnamefont{Michaluk}}, \bibnamefont{and}
  \bibinfo{author}{\bibfnamefont{A.}~\bibnamefont{Traxler}},
  \bibinfo{journal}{Phys.\ Rev.\ Phys.\ Educ.\ Res.}
  \textbf{\bibinfo{volume}{13}}, \bibinfo{pages}{020114}
  (\bibinfo{year}{2017}).

\bibitem[{\citenamefont{Karim et~al.}(2018)\citenamefont{Karim, Maries, and
  Singh}}]{kms18}
\bibinfo{author}{\bibfnamefont{N.~I.} \bibnamefont{Karim}},
  \bibinfo{author}{\bibfnamefont{A.}~\bibnamefont{Maries}}, \bibnamefont{and}
  \bibinfo{author}{\bibfnamefont{C.}~\bibnamefont{Singh}},
  \bibinfo{journal}{Eur.\ J.\ Phys.} \textbf{\bibinfo{volume}{39}},
  \bibinfo{pages}{025701} (\bibinfo{year}{2018}).

\bibitem[{\citenamefont{Lage et~al.}(2000)\citenamefont{Lage, Platt, and
  Treglia}}]{lpt00}
\bibinfo{author}{\bibfnamefont{M.~J.} \bibnamefont{Lage}},
  \bibinfo{author}{\bibfnamefont{G.~J.} \bibnamefont{Platt}}, \bibnamefont{and}
  \bibinfo{author}{\bibfnamefont{M.}~\bibnamefont{Treglia}},
  \bibinfo{journal}{J.\ Economic Educ.} \textbf{\bibinfo{volume}{31}},
  \bibinfo{pages}{30} (\bibinfo{year}{2000}).

\bibitem[{\citenamefont{Bishop and Verleger}(2013)}]{bv13}
\bibinfo{author}{\bibfnamefont{J.}~\bibnamefont{Bishop}} \bibnamefont{and}
  \bibinfo{author}{\bibfnamefont{M.~A.} \bibnamefont{Verleger}}, in
  \emph{\bibinfo{booktitle}{2013 ASEE Annual Conference \& Exposition}}
  (\bibinfo{year}{2013}),
  \urlprefix\url{https://www.asee.org/public/conferences/20/papers/6219/view}.

\bibitem[{ucs(2016, accessed 6/19/2017)}]{ucsddata}
\emph{\bibinfo{title}{First-time freshmen statistics}} (\bibinfo{year}{2016,
  accessed 6/19/2017}),
  \urlprefix\url{http://studentresearch.ucsd.edu/stats-data/admissions/freshmen.html}.

\bibitem[{\citenamefont{Stelzer et~al.}(2010)\citenamefont{Stelzer, Brookes,
  Gladding, and Mestre}}]{sbgm10}
\bibinfo{author}{\bibfnamefont{T.}~\bibnamefont{Stelzer}},
  \bibinfo{author}{\bibfnamefont{D.~T.} \bibnamefont{Brookes}},
  \bibinfo{author}{\bibfnamefont{G.}~\bibnamefont{Gladding}}, \bibnamefont{and}
  \bibinfo{author}{\bibfnamefont{J.~P.} \bibnamefont{Mestre}},
  \bibinfo{journal}{Am.\ J.\ Phys.} \textbf{\bibinfo{volume}{78}},
  \bibinfo{pages}{755} (\bibinfo{year}{2010}).

\bibitem[{\citenamefont{Sadaghiani}(2012)}]{sad12}
\bibinfo{author}{\bibfnamefont{H.~R.} \bibnamefont{Sadaghiani}},
  \bibinfo{journal}{Phys.\ Rev.\ ST Phys.\ Educ.\ Res.}
  \textbf{\bibinfo{volume}{8}}, \bibinfo{pages}{010103} (\bibinfo{year}{2012}).

\bibitem[{\citenamefont{Muller et~al.}(2008{\natexlab{a}})\citenamefont{Muller,
  Sharma, and Reimann}}]{msr08}
\bibinfo{author}{\bibfnamefont{D.~A.} \bibnamefont{Muller}},
  \bibinfo{author}{\bibfnamefont{M.~D.} \bibnamefont{Sharma}},
  \bibnamefont{and} \bibinfo{author}{\bibfnamefont{P.}~\bibnamefont{Reimann}},
  \bibinfo{journal}{Sci.\ Educ.} \textbf{\bibinfo{volume}{92}},
  \bibinfo{pages}{278} (\bibinfo{year}{2008}{\natexlab{a}}).

\bibitem[{\citenamefont{Muller et~al.}(2008{\natexlab{b}})\citenamefont{Muller,
  Bewes, Sharma, and Reimann}}]{mbsr08}
\bibinfo{author}{\bibfnamefont{D.}~\bibnamefont{Muller}},
  \bibinfo{author}{\bibfnamefont{J.}~\bibnamefont{Bewes}},
  \bibinfo{author}{\bibfnamefont{M.}~\bibnamefont{Sharma}}, \bibnamefont{and}
  \bibinfo{author}{\bibfnamefont{P.}~\bibnamefont{Reimann}},
  \bibinfo{journal}{J.\ Comp.\ Assisted Learning}
  \textbf{\bibinfo{volume}{24}}, \bibinfo{pages}{144}
  (\bibinfo{year}{2008}{\natexlab{b}}).

\bibitem[{\citenamefont{Mazur}(1997)}]{pimazur}
\bibinfo{author}{\bibfnamefont{E.}~\bibnamefont{Mazur}},
  \emph{\bibinfo{title}{Peer Instruction: A User's Manual}}
  (\bibinfo{publisher}{Prentice Hall Publishing}, \bibinfo{year}{1997}).

\bibitem[{\citenamefont{Knight}(2002)}]{fiveknight}
\bibinfo{author}{\bibfnamefont{R.~D.} \bibnamefont{Knight}},
  \emph{\bibinfo{title}{Five Easy Lessons: Strategies for Successful Physics
  Teaching}} (\bibinfo{publisher}{Pearson}, \bibinfo{year}{2002}).

\bibitem[{\citenamefont{Novak et~al.}(1999)\citenamefont{Novak, Patterson,
  Gavrin, and Christian}}]{jitt}
\bibinfo{author}{\bibfnamefont{G.}~\bibnamefont{Novak}},
  \bibinfo{author}{\bibfnamefont{E.}~\bibnamefont{Patterson}},
  \bibinfo{author}{\bibfnamefont{A.}~\bibnamefont{Gavrin}}, \bibnamefont{and}
  \bibinfo{author}{\bibfnamefont{W.}~\bibnamefont{Christian}},
  \emph{\bibinfo{title}{Just-in-Time Teaching: Blending Active Learning and Web
  Technology}} (\bibinfo{publisher}{Prentice Hall Publishing},
  \bibinfo{year}{1999}).

\bibitem[{\citenamefont{Crouch and Mazur}(2001)}]{cm01}
\bibinfo{author}{\bibfnamefont{C.~H.} \bibnamefont{Crouch}} \bibnamefont{and}
  \bibinfo{author}{\bibfnamefont{E.}~\bibnamefont{Mazur}},
  \bibinfo{journal}{Am.\ J.\ Phys.} \textbf{\bibinfo{volume}{69}},
  \bibinfo{pages}{970} (\bibinfo{year}{2001}).

\bibitem[{\citenamefont{Knight}(2016)}]{knightwkbk}
\bibinfo{author}{\bibfnamefont{R.~D.} \bibnamefont{Knight}},
  \emph{\bibinfo{title}{Student Workbook for Physics for Scientists and
  Engineers: A Strategic Approach with Modern Physics}}
  (\bibinfo{publisher}{Pearson}, \bibinfo{year}{2016}).

\bibitem[{\citenamefont{McDermott and Shaffer}(2001)}]{mcdermott}
\bibinfo{author}{\bibfnamefont{L.~C.} \bibnamefont{McDermott}}
  \bibnamefont{and} \bibinfo{author}{\bibfnamefont{P.~S.}
  \bibnamefont{Shaffer}}, \emph{\bibinfo{title}{Tutorials in Introductory
  Physics}} (\bibinfo{publisher}{Prentice Hall College Div},
  \bibinfo{year}{2001}).

\bibitem[{\citenamefont{Prather et~al.}(2013)\citenamefont{Prather, Slater,
  Adams, and Brissenden}}]{astro101}
\bibinfo{author}{\bibfnamefont{E.~E.} \bibnamefont{Prather}},
  \bibinfo{author}{\bibfnamefont{T.~F.} \bibnamefont{Slater}},
  \bibinfo{author}{\bibfnamefont{J.~P.} \bibnamefont{Adams}}, \bibnamefont{and}
  \bibinfo{author}{\bibfnamefont{G.}~\bibnamefont{Brissenden}},
  \emph{\bibinfo{title}{Lecture-Tutorials for Introductory Astronomy}}
  (\bibinfo{publisher}{Pearson}, \bibinfo{year}{2013}).

\bibitem[{\citenamefont{Prather et~al.}(2004)\citenamefont{Prather, Slater,
  Adams, Bailey, Jones, and Dostal}}]{prather04}
\bibinfo{author}{\bibfnamefont{E.~E.} \bibnamefont{Prather}},
  \bibinfo{author}{\bibfnamefont{T.~F.} \bibnamefont{Slater}},
  \bibinfo{author}{\bibfnamefont{J.~P.} \bibnamefont{Adams}},
  \bibinfo{author}{\bibfnamefont{J.~M.} \bibnamefont{Bailey}},
  \bibinfo{author}{\bibfnamefont{L.~V.} \bibnamefont{Jones}}, \bibnamefont{and}
  \bibinfo{author}{\bibfnamefont{J.~A.} \bibnamefont{Dostal}},
  \bibinfo{journal}{Astron.\ Educ.\ Rev.} \textbf{\bibinfo{volume}{3}},
  \bibinfo{pages}{122} (\bibinfo{year}{2004}).

\bibitem[{\citenamefont{Prather
  et~al.}(2009{\natexlab{a}})\citenamefont{Prather, Rudolph, and
  Brissenden}}]{prb09}
\bibinfo{author}{\bibfnamefont{E.~E.} \bibnamefont{Prather}},
  \bibinfo{author}{\bibfnamefont{A.~L.} \bibnamefont{Rudolph}},
  \bibnamefont{and}
  \bibinfo{author}{\bibfnamefont{G.}~\bibnamefont{Brissenden}},
  \bibinfo{journal}{Phys.\ Today} \textbf{\bibinfo{volume}{62}},
  \bibinfo{pages}{41} (\bibinfo{year}{2009}{\natexlab{a}}).

\bibitem[{\citenamefont{Prather et~al.}(2011)\citenamefont{Prather, Rudolph,
  and Brissenden}}]{prb11}
\bibinfo{author}{\bibfnamefont{E.~E.} \bibnamefont{Prather}},
  \bibinfo{author}{\bibfnamefont{A.~L.} \bibnamefont{Rudolph}},
  \bibnamefont{and}
  \bibinfo{author}{\bibfnamefont{G.}~\bibnamefont{Brissenden}},
  \bibinfo{journal}{Peer Rev.} \textbf{\bibinfo{volume}{13}},
  \bibinfo{pages}{27} (\bibinfo{year}{2011}).

\bibitem[{\citenamefont{Press et~al.}(2007)\citenamefont{Press, Teukolsky,
  Vetterling, and Flannery}}]{numrec}
\bibinfo{author}{\bibfnamefont{W.~H.} \bibnamefont{Press}},
  \bibinfo{author}{\bibfnamefont{S.~A.} \bibnamefont{Teukolsky}},
  \bibinfo{author}{\bibfnamefont{W.~T.} \bibnamefont{Vetterling}},
  \bibnamefont{and} \bibinfo{author}{\bibfnamefont{B.~P.}
  \bibnamefont{Flannery}}, \emph{\bibinfo{title}{Numerical Recipes: The Art of
  Scientific Computing, 3rd ed.}} (\bibinfo{publisher}{Cambridge University
  Press}, \bibinfo{year}{2007}).

\bibitem[{\citenamefont{Bransford et~al.}(2000)\citenamefont{Bransford, Brown,
  and Cocking}}]{learn}
\bibinfo{editor}{\bibfnamefont{J.~D.} \bibnamefont{Bransford}},
  \bibinfo{editor}{\bibfnamefont{A.~L.} \bibnamefont{Brown}}, \bibnamefont{and}
  \bibinfo{editor}{\bibfnamefont{R.~R.} \bibnamefont{Cocking}}, eds.,
  \emph{\bibinfo{title}{How People Learn: Brain, Mind, Experience and School}}
  (\bibinfo{publisher}{National Academies Press}, \bibinfo{year}{2000}).

\bibitem[{\citenamefont{Prather
  et~al.}(2009{\natexlab{b}})\citenamefont{Prather, Rudolph, Brissenden, and
  Schlingman}}]{lsci}
\bibinfo{author}{\bibfnamefont{E.~E.} \bibnamefont{Prather}},
  \bibinfo{author}{\bibfnamefont{A.~L.} \bibnamefont{Rudolph}},
  \bibinfo{author}{\bibfnamefont{G.}~\bibnamefont{Brissenden}},
  \bibnamefont{and} \bibinfo{author}{\bibfnamefont{W.~M.}
  \bibnamefont{Schlingman}}, \bibinfo{journal}{Am.\ J.\ Phys.}
  \textbf{\bibinfo{volume}{77}}, \bibinfo{pages}{320}
  (\bibinfo{year}{2009}{\natexlab{b}}).

\bibitem[{\citenamefont{Grunspan et~al.}(2016)\citenamefont{Grunspan, Eddy,
  Brownell, Wiggins, Crowe, and Goodreau}}]{gebwcg}
\bibinfo{author}{\bibfnamefont{D.~Z.} \bibnamefont{Grunspan}},
  \bibinfo{author}{\bibfnamefont{S.~L.} \bibnamefont{Eddy}},
  \bibinfo{author}{\bibfnamefont{S.~E.} \bibnamefont{Brownell}},
  \bibinfo{author}{\bibfnamefont{B.~L.} \bibnamefont{Wiggins}},
  \bibinfo{author}{\bibfnamefont{A.~J.} \bibnamefont{Crowe}}, \bibnamefont{and}
  \bibinfo{author}{\bibfnamefont{S.~M.} \bibnamefont{Goodreau}},
  \bibinfo{journal}{PLOS\ ONE} \textbf{\bibinfo{volume}{11}},
  \bibinfo{pages}{1} (\bibinfo{year}{2016}).

\bibitem[{\citenamefont{Nosek et~al.}(2002)\citenamefont{Nosek, Banaji, and
  Greenwald}}]{nbg02}
\bibinfo{author}{\bibfnamefont{B.~A.} \bibnamefont{Nosek}},
  \bibinfo{author}{\bibfnamefont{M.~R.} \bibnamefont{Banaji}},
  \bibnamefont{and} \bibinfo{author}{\bibfnamefont{A.~G.}
  \bibnamefont{Greenwald}}, \bibinfo{journal}{Group Dynamics: Theory, Research,
  and Practice} \textbf{\bibinfo{volume}{6}}, \bibinfo{pages}{101}
  (\bibinfo{year}{2002}).

\bibitem[{\citenamefont{Nosek and Smyth}(2011)}]{ns11}
\bibinfo{author}{\bibfnamefont{B.~A.} \bibnamefont{Nosek}} \bibnamefont{and}
  \bibinfo{author}{\bibfnamefont{F.~L.} \bibnamefont{Smyth}},
  \bibinfo{journal}{Am.\ Educ.\ Res.\ J.} \textbf{\bibinfo{volume}{48}},
  \bibinfo{pages}{1125} (\bibinfo{year}{2011}).

\end{thebibliography}
\end{document}